\documentclass[12pt,titlepage]{article}
\pdfoutput=1

\usepackage[utf8]{inputenc} 
\usepackage[letterpaper,margin=1in]{geometry}
\usepackage{graphicx} 
\usepackage{setspace}
\usepackage{changepage}
\usepackage{float}
\usepackage{amsmath}
\usepackage[sort&compress]{natbib}
\usepackage[margin=10pt,font=small,labelfont=bf]{caption}
\usepackage[affil-it]{authblk}
\renewcommand{\thefootnote}{\fnsymbol{footnote}}

\title{Singlet fission for quantum information and quantum computing: The parallel \emph{JDE} model} 
\author{Kori Smyser}
\author{Joel Eaves\thanks{Email: joel.eaves@colorado.edu}}
\affil{Department of Chemistry, University of Colorado. Boulder, Colorado 80309-0215\\ United States}

\begin{document}
\maketitle
\renewcommand{\thefootnote}{\arabic{footnote}}
\begingroup
\centering\small
\textbf{Abstract}\par\medskip
\begin{adjustwidth}{1cm}{2cm}
Singlet fission is a photoconversion process that generates a doubly excited, maximally spin entangled  pair state. This state has applications to quantum information and computing that are only beginning to be realized. In this article, we construct and analyze a spin-exciton hamiltonian to describe the dynamics of the two-triplet state. We find the selection rules that connect the doubly excited, spin-singlet state to the manifold of quintet states and comment on the mechanism and conditions for the transition into formally independent triplets.  For adjacent dimers that are oriented and immobilized in an inert host, singlet fission can be strongly state-selective.  We make predictions for electron paramagnetic resonance experiments and re-assign some transitions from recent literature.  Our results give conditions for which magnetic resonance pulses can drive transitions between optically polarized magnetic sublevels of the two-exciton states, making it possible to realize quantum gates at room temperature in these systems.
\end{adjustwidth}

\par\endgroup
\bigskip\noindent

\section*{Introduction}
Materials capable of storing and manipulating quantum data must maintain quantum coherences and entanglement over timescales that are orders of magnitude longer than the system's quantum beat period \cite{DiVincenzo2000}. But quantum states are fragile, and most materials do not sustain quantum coherences when temperatures are in excess of a few Kelvin. This ``tyranny of low temperature" is a major hurdle to realizing accessible quantum computing and information technologies.

From a quantum information perspective, the photoproducts of singlet fission can solve two outstanding problems associated with the tyranny of low temperature, and they do it in complementary ways. Each corresponds to one of the  criteria for quantum computing set forth by DiVincenzo \cite{DiVincenzo2000}. To paraphrase: a system capable of quantum information processing must be prepared in a \emph{pure quantum state}, not a mixed one\cite{Warren1997}. Once initialized, one must be able to execute a deterministic sequence of unitary operations, or quantum gates, on that state so that it can be coaxed into collapsing on the final state, which is the solution to a computational problem, with high probability and in polynomial time\cite{Shor1999,DiVincenzo1999}. 

\begin{figure}
    \centering
    \includegraphics{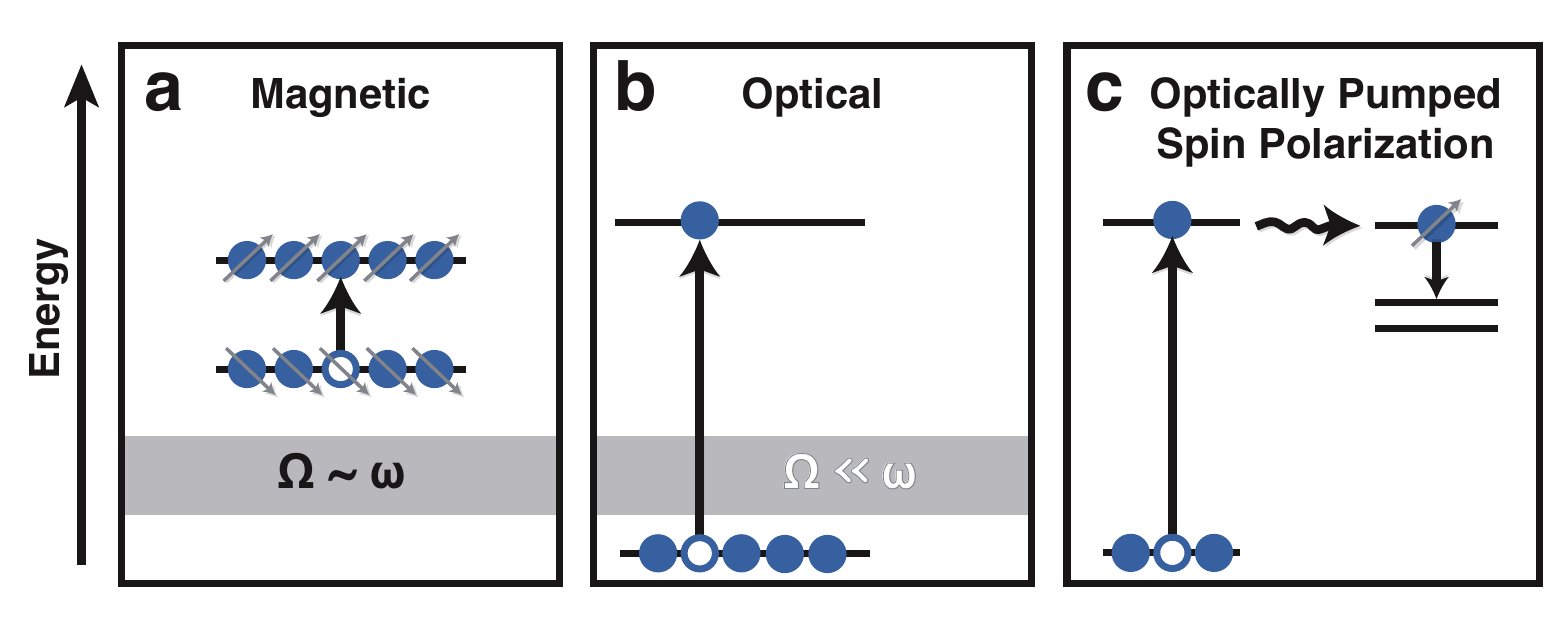}
    \caption{Optically induced spin polarization generates states for quantum information applications near room temperature. \textbf{a} In magnetic resonance experiments, the Rabi frequency, $\Omega$, can be comparable to the transition (Larmor) frequency, $\omega$. In this strong field limit, it is possible to completely manipulate a quantum state (qubit). However, the energy gap between states is small relative to $k_BT$. This is a source of uncertainty when the state is initialized. \textbf{b} In optical transitions, the energy gap is large relative to $k_BT$. But the Rabi frequency is much less than the transition frequency, which means that gate operations done with weak optical fields will be incomplete and noisy. \textbf{c} By coupling optical excitations to an internal conversion process, such as singlet fission (wavy arrow), one may capitalize on the advantages of both methods, provided that the relaxation is state-selective.}
    \label{fgr:FIG1}
\end{figure}

Because magnetic resonance experiments operate in the ``strong-field" regime where the Rabi frequency, $\Omega$, is comparable to the transition frequency, $\omega$ (Fig.~\ref{fgr:FIG1}a), they can drive arbitrary unitary operations between quantum spin states \cite{2001Vandersypen}. But the gap between the ground and excited states in these experiments is small relative to $k_B T$, and as a result, there is a great deal of thermally generated uncertainty in the initial state of the system. In the language of the density matrix, the initial state is mixed, not pure. Optical experiments, by contrast, have a large gap between ground and excited states relative to $k_B T$ (Fig.~\ref{fgr:FIG1}b), so that thermal fluctuations do not generate appreciable uncertainty in the ensemble of initial states. The Rabi frequency in optical experiments, however, is perturbatively small relative to the transition frequency. It is therefore very difficult or impossible to induce a population inversion in an optical experiment, which severely limits the ability of a purely optical experiment to perform quantum gate operations.

In more recent years, researchers have become interested in systems where it is possible to generate spin polarization through optical pumping (Fig.~\ref{fgr:FIG1}c) \cite{Fataftah2018,Gaita-Arino2019}. The optical field removes the uncertainty in the initial state of the system, and magnetic resonance experiments on the optically prepared photoproduct, well in the strong field regime, perform the quantum gate operations. The nitrogen-vacancy (NV) center in diamond is an example of a system that operates on these principles. In solid-state systems, like the NV center, spin centers are implanted into the material post-synthesis. Because the defects are often randomly dispersed, these materials have problems with scalability that might be overcome in molecular systems that are synthesized from the bottom up.

In this article, we explore the phenomenon of singlet fission as a novel platform on which one might build quantum data structures and gates near room temperature. Singlet fission is a photophysical interconversion process that takes place between specifically designed organic chromophores where, following excitation with a photon ($\gamma$), an optically bright singlet state, $S_0S_1$, rapidly relaxes into a doubly excited, spin-singlet state, $^1\text{TT}$ (equations~\ref{eq:absorption}-\ref{eq:sf}), \cite{Smith2010,Smith2013,Pope1969}

\begin{eqnarray}
\label{eq:absorption}
\gamma + S_0S_0 & \rightleftharpoons & S_0S_1, \\
\label{eq:sf}
S_0S_1& \rightleftharpoons & ^1 \text{TT}.
\end{eqnarray}

In this notation, the superscript indicates the state's multiplicity in terms of the total spin $S$, and the $\text{TT}$ designates the spatial nature of the spin wavefunction. The $|^1\text{TT}\rangle$ spin wavefunction is a maximally entangled, coherent superposition over three of the nine spin sublevels belonging to the two triplets on the chromophores\cite{Burdett2012, Scholes2015, Bardeen2019}. In the last few years, electron paramagnetic resonance (EPR) experiments have combined pulsed optical laser excitation with magnetic resonance to reveal that the  $^1\text{TT}$ state evolves into various $^{2S+1}\text{TT}_M$ exciton states \cite{Lubert-Perquel2018, Weiss2016, Nagashima2018, Tayebjee2016, Matsui2019, Chen2019, Basel2018, Basel2017, Sakai2018}

\begin{equation}
\label{eq:sf_non1}
^1 \text{TT} \rightleftharpoons {^{2S+1}\text{TT}}_M,
\end{equation}
where the subscript, $M$, refers to the magnetic sublevel $-S\le M\le +S$. Although singlet fission is a spin-conserving process (equation~\ref{eq:sf}), the triplet pair states (equation~\ref{eq:sf_non1}) are not eigenfunctions of the electronic spin hamiltonian; they are non-stationary and evolve in time. 

In crystalline systems, these excitons may hop to neighboring sites, becoming increasingly more distant, and eventually unpair into $\text{T+T}$ \cite{Swenberg1973}:

\begin{equation}
\label{eq:sf_non2}
^{2S+1}\text{TT}_M \rightleftharpoons \text{T+T}.
\end{equation}
There has not been a consistent microscopic theory that can explain the set of relaxation phenomena embodied in equations 1-4.

While the lion's share of attention over the past decade has focused on maximizing the conversion of the $^1\text{TT}$ state to $\text{T+T}$ for solar energy applications\cite{Hanna2006,Paci2006}, we argue that for quantum information applications, one should seek to instead \emph{limit} the decay into independent excitons, by designing molecules that make the conversion from $^1\text{TT}$ to $^{2S+1}\text{TT}_M$ as state-selective as possible.

In this work we consider singlet fission between chromophores in two classes of systems that are widespread in the literature: covalently linked organic dimer molecules\cite{Gilligan2019} and organic crystals comprised of chromophore pairs\cite{Lubert-Perquel2018}.  As is often the case, the selection rules governing the quantum relaxation phenomena depend sensitively on molecular symmetries. In particular, a pair of identical chromophores, where one molecule is related to the other by a translation (Fig.~\ref{fgr:FIG2}b), will also exhibit a permutation symmetry for the exciton triplet pair. As we show, that symmetry isolates the  $^3\text{TT}$ triplet states so that spin relaxation only proceeds between $^1\text{TT}$ and $^5\text{TT}$ states. The most state-selective relaxation occurs between the $^1\text{TT}$ singlet state and the $^5\text{TT}_{0}$ quintet state in an ordered and immobilized system of molecular dimers that have their magnetic principal axes mutually parallel to one another.

\section*{Results} 

\subsection*{The Spin-Exciton Hamiltonian} 

We follow with a derivation of the spin-exciton hamiltonian for the triplet pairs in equation~\ref{eq:sf_non1}, exploiting approximations for the light atom molecules characteristic of singlet fission chromophores. The singlet fission process depicted in equation~\ref{eq:sf} is often much faster (picosecond or sub-picosecond) than the timescales on which the states in equation~\ref{eq:sf_non2} evolve (nanoseconds to microseconds). The spin-orbit interaction is small and is often ignored, but it is straightforward to include\cite{Pedash2002}. 

\begin{figure}
    \centering
    \includegraphics{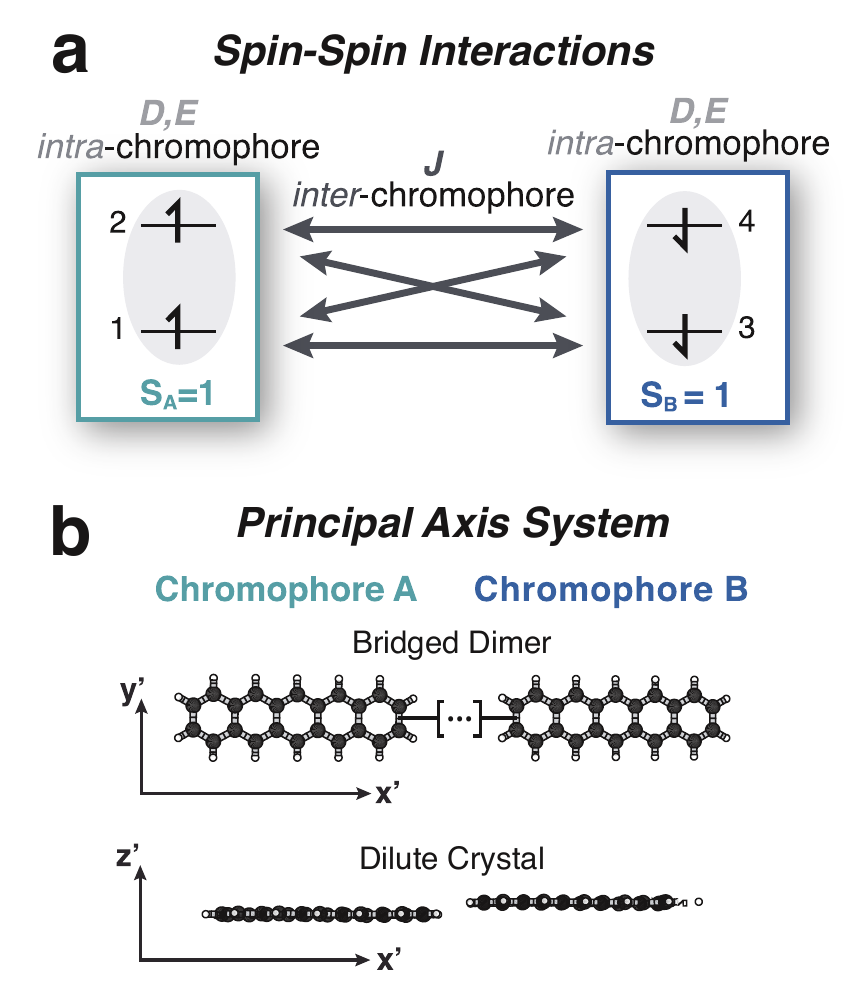}
    \caption{The \emph{JDE} model. \textbf{a} The products of singlet fission are doubly excited states, with one electron in each frontier molecular orbital (numbered 1-4) of each chromophore (labelled $A$ or $B$). The spin of each chromophore is a triplet so that $S_A=S_B=1$. We assume that all \emph{inter}-chromophore isotropic exchange interactions $J$ (double-headed arrows) are equivalent. The \emph{intra}-chromophore interaction is spin-dipole in origin and is characterized by the axial and rhombic EPR parameters, $D$ and $E$, respectively. In the parallel \emph{JDE} model, we assume that $D=D_A=D_B$ and $E=E_A=E_B$. This figure is similar in spirit to a picture presented in a recent review, though the nature of the spin-spin interactions there differs from what we present here\protect{\cite{Miyata2019}}. \textbf{b} The parallel \emph{JDE} model describes two translationally invariant chromophores that have all zero-field splitting principal axes $\left(x',y',z'\right)$ parallel to one another. These chromophores may be covalently linked (top) or doped into a photophysically inactive host matrix (bottom).}
    \label{fgr:FIG2}
\end{figure}

After making the Born-Oppenheimer approximation, suppressing orbital degrees of freedom and ignoring hyperfine interactions, we begin with a general hamiltonian that is bilinear in all spin-spin interactions,
\begin{equation}
\label{eq:All-Electron}
    {\cal H} = \frac{1}{2}\sum^4_{\substack{i,j=1 \\ i\neq j}} \mathbf{s}_i\cdot\mathbf{O}_{ij}\cdot\mathbf{s}_j,
\end{equation}
where $\mathbf{s}_i$ is the electron spin operator for orbital $i$, and the indices $i$ and $j$ enumerate the HOMO and LUMO levels in a frontier molecular orbital description of the chromophore pair (Fig.~\ref{fgr:FIG2}a). Similarly, the $\mathbf{g}$-tensors in singlet fission chromophores  are often isotropic\cite{Smith2010}, though anisotropy in the $\mathbf{g}$-tensors can also be included perturbatively. $\mathbf{O}_{ij}$ is a rank-2 tensor that accounts for the spin-spin interactions between the four electrons in the four orbitals. Within this framework, $\mathbf{O}_{ij}$ depends implicitly on integrals over spatial wavefunctions. Akin to electronically nonadiabatic effects in molecules, fluctuations in nuclear coordinates and exciton hopping can make the $\mathbf{O}_{ij}$ time-dependent \cite{Collins2019,Swenberg1973}. These time-dependent $\mathbf{O}_{ij}$ parameters can drive transitions between different spin states of the spin-exciton hamiltonian.

The interaction tensor $\mathbf{O}_{ij}$ for each electron pair can be decomposed into three separate terms: a scalar '``isotropic'' part, an antisymmetric tensor of rank one, and a traceless, anisotropic tensor of rank two. The isotropic term yields the usual Dirac-Heisenberg exchange coupling\cite{Weil2007}. Unlike the isotropic exchange interaction, both the antisymmetric and the anisotropic terms are formally relativistic in nature, and in light-atom molecules they are much smaller than the isotropic term \cite{Bencini1990}. For example, the isotropic exchange interaction in a single pentacene molecule splits the singlet and triplet levels by about 1 eV\cite{Smith2010} while the anisotropic interaction splits the triplet levels by about 1 GHz or $\sim 10^{-6}$ eV\cite{VanStrien1980}. The antisymmetric term is usually negligible in aromatic hydrocarbons and so we ignore it.

To simplify the hamiltonian, we decompose it into intra-chromophore and inter-chromo\-phore interactions. Because we are not interested in modeling, for example, the transitions between the singlet and triplet state on a single chromophore, we do not include the intra-chromophore isotropic exchange interaction in the hamiltonian. We do, however, keep the \emph{intra}-chromophore anisotropic coupling. We also include the isotropic \emph{inter}-chromophore exchange interaction but discard the much smaller inter-chromophore anisotropic coupling. Finally, we set all of the inter-chromophore isotropic exchange interactions equal to the same number, $J$. The spin-exciton hamiltonian then takes a compact form,
\begin{equation}
\label{eq:exciton}
{\cal H} = J\;\mathbf{S}_A\cdot\mathbf{S}_B + H_A + H_B,
\end{equation}
where $\mathbf{S}_A = \mathbf{s}_1 + \mathbf{s}_2$ and $\mathbf{S}_B = \mathbf{s}_3 + \mathbf{s}_4$ are the spin operators associated with chromophores $A$ and $B$. 

The first term in equation~\ref{eq:exciton} is the isotropic exchange interaction between two triplet excitons.  $H_A = D\left(S_{Az'}^2 - \mathbf{S}_A^2/3\right) + E\left(S_{Ax'}^2 - S_{Ay'}^2\right) $ and $H_B=D\left(S_{Bz'}^2 - \mathbf{S}_B^2/3\right) + E\left(S_{Bx'}^2 - S_{By'}^2\right)$ are the hamiltonians associated with the intra-chromophore anisotropic interactions, taken to be of the spin-dipole form, written in the canonical ``zero-field splitting" (ZFS) form from EPR literature\cite{Weil2007}. The primed coordinates denote the principal axes of the magnetic dipole tensor for chromophore $A$ and chromophore $B$. The highest symmetry case is the one we analyze here, where all principal directions for chromophore $A$ are parallel to those of chromophore $B$.

In small organic chromophores like pentacene and tetracene, the single quantum triplet excited state is strongly localized  \cite{Cudazzo2012}. The triplet states, each bound to a single chromophore, form molecular (Frenkel) excitons\cite{Scholes2006}. Like the spatial wavefunctions, the single quantum spin wavefunctions should also be tightly bound and have a definite triplet multiplicity. Equation~\ref{eq:exciton} describes the interactions between two triplet Frenkel spin-excitons, modeled as two spin-1 objects, each one interacting with itself through the spin-dipole interaction and coupled to one another through exchange.

Because the hamiltonian depends only on the parameters $J$, $D$ and $E$ and the principal axes of $A$ and $B$ are parallel, we refer to the spin-exciton hamiltonian in equation \ref{eq:exciton} as the parallel \emph{JDE} model. As we will show, restricting the axes of $A$ and $B$ to be parallel imposes symmetries that make relaxation between magnetic sublevels maximally state-selective. All of the parameters in the \emph{JDE} model can either be computed in electronic structure or measured in mixed optical/magnetic resonance experiments.
    
While hamiltonians similar to equation~\ref{eq:exciton} have appeared more recently in literature, researchers use approximations that make equation~\ref{eq:exciton} spin-conserving \cite{Weiss2016,Chen2019,Collins2019,Matsuda2020,Tayebjee2016,Lubert-Perquel2018,Wakasa2015}. Methods of arriving at a spin-conserving \emph{JDE} model differ between authors, but they amount to making an effective spin approximation, such that, for example, $ S_{Az}^2 + S_{Bz}^2 \approx S_z^2$. One then discards the isotropic exchange term, which is spin-conserving anyway, and writes the hamiltonian with renormalized $D$ and $E$ parameters, where the $A$ and $B$ site-spin operators are replaced by total spin operators. The $D$ and $E$ zero-field splitting parameters in crystals are renormalized to the spatially averaged $D^*$ and $E^*$ \cite{Yarmus1972,Sternlicht1961}. In the ``strong exchange limit" $|J|/|D| \gg 1$, $D$ becomes $D/3$, but only for the quintets\cite{Weiss2016}. These approximations are suitable for work that does not consider transitions between states of different multiplicity, but they are manifestly incapable of describing the kind of intersystem crossing, from $^1\text{TT}$ to $^5\text{TT}_M$ for example, that recent EPR experiments have observed. These newer measurements necessitate the development of the theory presented here.

\subsubsection*{Spin Dynamics and Selection Rules at Zero Applied Field} 

The exchange term $J\mathbf{S}_A\cdot \mathbf{S}_B$ is the largest energy scale in the hamiltonian. It is rotationally invariant and diagonal in the total spin representation. We therefore take the diagonal elements of the \emph{JDE} hamiltonian in the basis of total $\mathbf{S}^2$ and $S_z$ as our reference hamiltonian, $H_0$. The couplings between states of different multiplicity are off-diagonal perturbations, $V$, that cause transitions between these levels. They depend solely on $D$ and $E$. The terms that comprise $V$ are bilinear products of $\mathbf{S}_A$ and $\mathbf{S}_B$ Cartesian spin operators, and in the parallel case the largest of these terms is $D S_{Az}S_{Bz}$. It has the form of an effective ``Zeeman" interaction, where the $z$-component of the magnetic field produced by exciton spin-$A$ couples to the magnetic dipole of spin-$B$.

\begin{figure}
    \centering
    \includegraphics{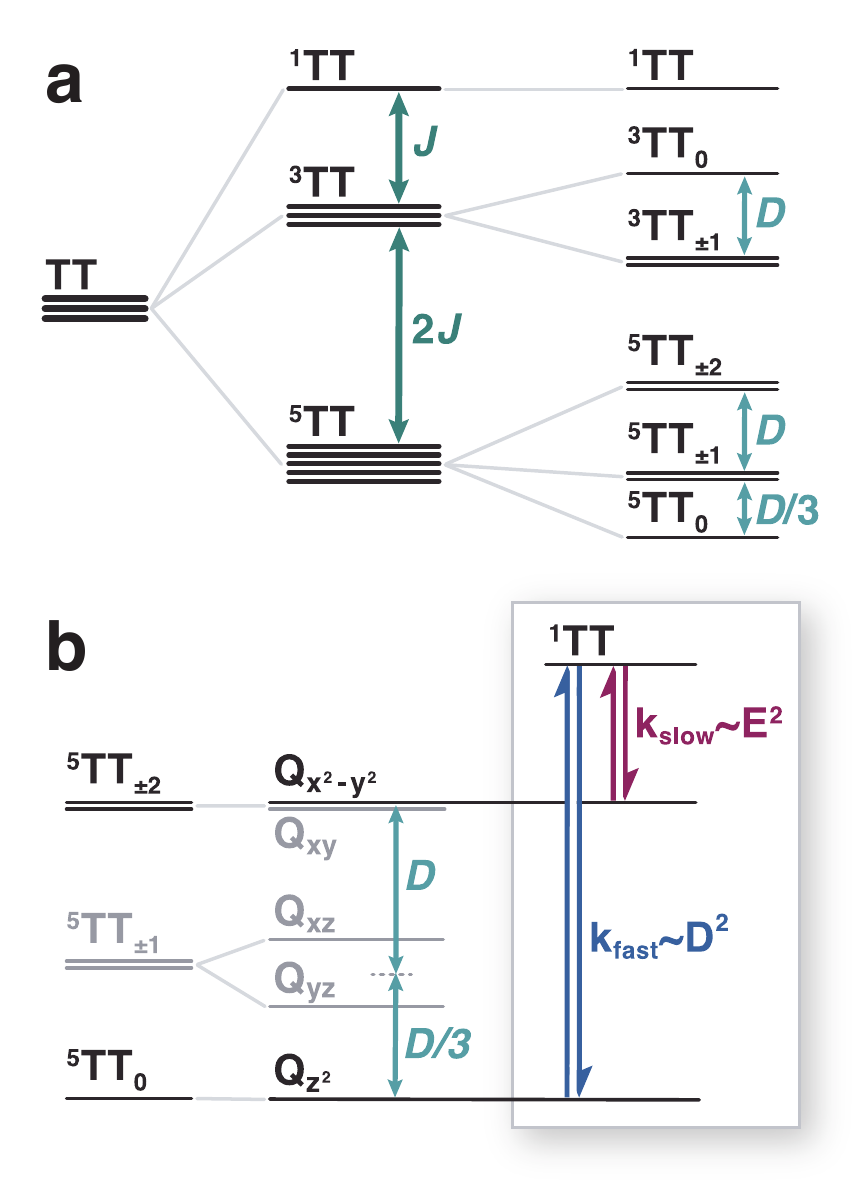}
    \caption{Zero-field energy and selection rules. \textbf{a} Energy-level correlation diagram of the \emph{JDE} model in zero applied field. Interaction energies and splittings decrease in magnitude going from left to right. The sign of $J$ orders the states of total $S$; we choose $J<0$ in analogy to  Hund's rule. The choice $D>0$ corresponds to literature values for pentacene\protect{\cite{Swenberg1973}}. $D$ is the second largest energy scale and it lifts the degeneracy of states with different magnitudes of the total spin projection quantum number $|M|=0,1,2$. \textbf{b} The $Q$ states are linear combinations of degenerate pairs of $^5\text{TT}_{\pm M}$ states with Cartesian subscripts that indicate analogies to the d-orbitals. Transitions from $^1\text{TT}$ are only allowed to two $Q$ states (black). The $Q_{z^2}$ state may be populated through a non-adiabatically fast interconversion process which goes as $k_{\text{fast}}\sim D^{2}$. A much slower process $k_\text{slow}\sim E^{2}$ allows transition from the $^1\text{TT}$ state to the $Q_{x^2-y^2}$ state.}
    \label{fgr:FIG3}
\end{figure}

The energy level diagram for the diagonal states of the reference hamiltonian, $H_0$, appears in Fig.~\ref{fgr:FIG3}a. States of different multiplicity are split by the large inter-chromophore exchange interaction, $J$. The singlet-quintet splitting between $^1\text{TT}$ and $^5\text{TT}$ is three times the singlet-triplet splitting. The much smaller intra-chromophore axial ZFS interaction, $D$, splits magnetic, $M$, sublevels within each manifold. The rhombicity parameter, $E$ --- the smallest energy scale in the hamiltonian --- gives rise to fine structure splittings between $M$-sublevels of $H_0$.

When evaluating the hamiltonian, we express it in terms of spherical tensor operators and apply the Wigner-Eckhart theorem \cite{Edmonds1996}. In our evaluation, the ``renormalization" of $D$, for example, to $D/3$ for quintets (Fig.~\ref{fgr:FIG3}a),  is a direct consequence of the Wigner-Eckart theorem. We assume that the ZFS, $D$ and $E$ parameters are independent of nuclear coordinates and that the time-dependent nuclear motion appears in $J$. A recent paper that numerically simulates spin dynamics in singlet fission by Chen  et  al. makes a similar approximation \cite{Chen2019}.

Equation~\ref{eq:exciton} allows transitions from $^1\text{TT}$ to $^5\text{TT}_{0}$ and $^5\text{TT}_{\pm2}$. Forming linear combinations from the sums and differences of each pair of degenerate states breaks the degeneracy of the $^5\text{TT}_{\pm 1}$ states by $2|E|$ (Fig.~\ref{fgr:FIG3}b). These linear combinations of the $M$-sublevels are analogous to those of the $\ell = 2$ spherical harmonics that are taken to construct the d-orbitals\cite{Benk1981}, and so we label the states accordingly (Fig.~\ref{fgr:FIG3}b). In this representation, the $|^1\text{TT}\rangle$ state only couples to $|Q_{x^2 - y^2}\rangle =\left(|{^5\text{TT}_{2}}\rangle + |{^5\text{TT}_{-2}}\rangle\right)/\sqrt{2}$ and to $|{Q_{z^2}}\rangle = |{^5\text{TT}_{0}}\rangle$. These states are not simply the eigenstates of $H_{ZFS}$. The state-selectivity follows from the symmetries of how the Cartesian $Q$ states, rather than the $|S,M\rangle$ states, transform under the rotation operations that characterize the symmetries of the zero-field hamiltonian.

In a rate theory for relaxation, the relaxation rate from $^1\text{TT}$ to $Q_{z^2}$ is proportional to $D^2$ and the rate from $^1\text{TT}$ to $Q_{x^2 - y^2}$ is proportional to $E^2$.  Because $|D| \gg |E|$, the dominant relaxation channel is from $^1\text{TT}$ to $Q_{z^2}$  (Fig.~\ref{fgr:FIG3}b). These selection rules are strict for identical chromophores with parallel symmetry, but upon breaking this symmetry, transitions from the quintet $^5\text{TT}$ to the triplet manifold $^3\text{TT}$ become allowed while transitions from the $^1\text{TT}$ to the $^3\text{TT}$ manifold remain forbidden. In the absence of parallel symmetry, transitions also become allowed between $^1\text{TT}$ and all of the quintet sublevels. 

Equation~\ref{eq:exciton} can describe how the unpairing of $^1\text{TT}$ to independent triplets $T+T$ takes place (equation~\ref{eq:sf_non2}) \cite{Collins2019}. By measuring quantum beats in delayed fluorescence spectra, Burdett and Bardeen showed, rather convincingly, that exciton unpairing can occur in crystalline tetracene \cite{Burdett2012}. In a crystal, the exciton hopping rate is fast. As a result, $J$, which depends sensitively on the distance between chromophores, will rapidly go to zero. This scenario can be modeled using a quantum quench with the time-dependent hamiltonian ${\cal H}(t) = H_A + H_B + (1 - \theta(t))J\mathbf{S}_A\cdot \mathbf{S}_B$, where $\theta(t)$ is the Heaviside step function, and the quench occurs for $t>0$. After the initial transients associated with decoherence and population relaxation phenomena subside, detailed balance shows that the system's reduced density matrix ($\rho$) will factorize into a thermal product state. Because $H_A$ and $H_B$ commute, at times that are long compared to the spin relaxation times the density matrix becomes $\rho \sim \exp\left(-\beta H \right)/Z=\exp\left(-\beta H_A\right)\exp\left(-\beta H_B\right)/Z_AZ_B = \rho_A \otimes \rho_B$. The two triplet exciton states become formally independent. 

Exciton entanglement diminishes as $J$ becomes smaller. This implies that dimers will preserve entanglement on longer timescales than crystals with mobile excitons. In tetracene crystals, the loss of coherence has been observed to occur on timescales of tens of nanoseconds, which is not much longer than the quantum gate switching times given by the inverse characteristic EPR transition frequencies\cite{Burdett2012}. Once thermalized, these two triplets offer no quantum advantage over single triplets prepared through more standard intersystem crossing processes. 

For quantum information applications, one should focus attention on dimers where chromophores are covalently bound or packed together as a minority component in a crystal so that the exciton hopping rate to the host is negligible \cite{Lubert-Perquel2018}. We refer to the latter category as a ``dilute crystal." In these systems, $J$ may fluctuate about a nonzero value, but it cannot go to zero.  We impose the condition that $|J| \gg |D|$. For organic chromophores, such as polyacenes, it is also often the case that $|D| \gg |E|$.

In dimers, transitions can occur from rare fluctuations of the bare energy gaps between states of total $|{S,M}\rangle$. These fluctuations are driven by nuclear motions, and we assume that the energy gap embodied in a time-dependent $J$ obeys Gaussian statistics. This scenario is valid so long as $|J| \gg |D|$ and the energy gap obeys linear response with respect to the nuclear motions. The resulting theory is completely analogous to Marcus' theory of nonadiabatic electron transfer and the F\"orster-Dexter theory of exciton hopping, where the ZFS parameters play the role of the nonadiabatic coupling matrix elements. Transitions between initial and final states take the form $k_{i \rightarrow f} = F |\langle i|V|f\rangle|^2$, where $F$ is a Franck-Condon weighted density of states. In principle, $F$ incorporates a thermal factor between the various final states that is the result of summing over nuclear fluctuations. Detailed balance, however, gives the condition that $k_{i\rightarrow f}/k_{f\rightarrow i}=\exp(-\beta\left(E_i - E_f\right))$. Given that the various quintet states are split by about 0.05 $cm^{-1}$, ignoring the temperature dependence of the prefactor is a safe approximation for temperatures above about 1 K. It is a straightforward matter to extend this analysis to a case where $|J|<|D|$, by first diagonalizing the \emph{JDE} hamiltonian and then applying second-order perturbation theory in the exciton-heat bath coupling. This approach would resonate closely with the much earlier work of Johnson and Merrifield on delayed fluorescence in molecular crystals\cite{Johnson1970} that has since been applied to recent experiments \cite{Bardeen2019}.

\subsection*{EPR Spectroscopy}

Recent experiments have employed an optical pump/EPR probe scheme to observe the fate of exciton polarization following singlet fission \cite{Basel2017,Basel2018,Chen2019,Collins2019,Matsui2019,Nagashima2018,Sakai2018,Tayebjee2016,Weiss2016,Matsuda2020}. Many of these experiments use field-swept EPR as the probe, where the system is subjected to a static magnetic field, $\mathbf{B_0}$, along the laboratory $z$-axis. The static field splits the magnetic sublevels while an oscillatory microwave field, $\mathbf{B_1}$, polarized in the $xy$-plane, induces transitions between them. In these experiments, one finds resonances as a function of the static field strength, $B_0$.

Many experiments use X-band EPR (8-12 GHz) and for small organic chromophores this is in the strong field limit, where states are split by much more than $D$. To model these experiments, we introduce the Zeeman term, $H_{Zeeman} = g\mu_B B_0 S_z$, into equation~\ref{eq:exciton} and choose the quantization axis along the lab, or Zeeman, $z$-axis. The static Zeeman field splits states of different $M$ but not states of different $S$; the $^1\text{TT}$ state is unaffected by the Zeeman field.

Because the Zeeman field induces splittings that are large compared to those of $H_{ZFS}$, the reference hamiltonian, $H_0$, changes. To construct it, we first project out the quintet block to find its eigenstates, $|\alpha\rangle = \sum_{M} c_{M,\alpha}|^5\text{TT}_M\rangle$. These states, the adiabats (Fig.~\ref{fgr:FIG5}b), adiabatically follow $B_0$. The hamiltonian is then re-expressed in the adiabatic basis, with the reference hamiltonian, $H_0 = \sum_\alpha |{\alpha}\rangle \epsilon_\alpha \langle{\alpha}|$, and coupling to $|{^1\text{TT}}\rangle$ defined accordingly. Transitions occur between the adiabatic sublevels and have a spectrum given by the Golden Rule $I = \sum_{\alpha,\beta} |\langle{\alpha}|S_x |{\beta}\rangle|^2(P_{\alpha} - P_{\beta})\delta(\epsilon_\alpha - \epsilon_\beta)$ \cite{Mabbs1992}, where $P_\alpha = Tr(\rho |{\alpha}\rangle\langle{\alpha}|)$ is the population in state $|{\alpha}\rangle$. While the time-dependence of the populations can be, and has been, measured, we focus attention on the ``prompt" EPR spectrum that interrogates the initial population of the exciton magnetic sublevels immediately following singlet fission, where the short-time approximation $P_{\alpha} \sim |\langle{^1\text{TT}}| H|{\alpha}\rangle|^2$ is valid. With these provisions in place, there are no adjustable parameters for the calculated EPR spectra.

\begin{figure}
    \centering
    \includegraphics{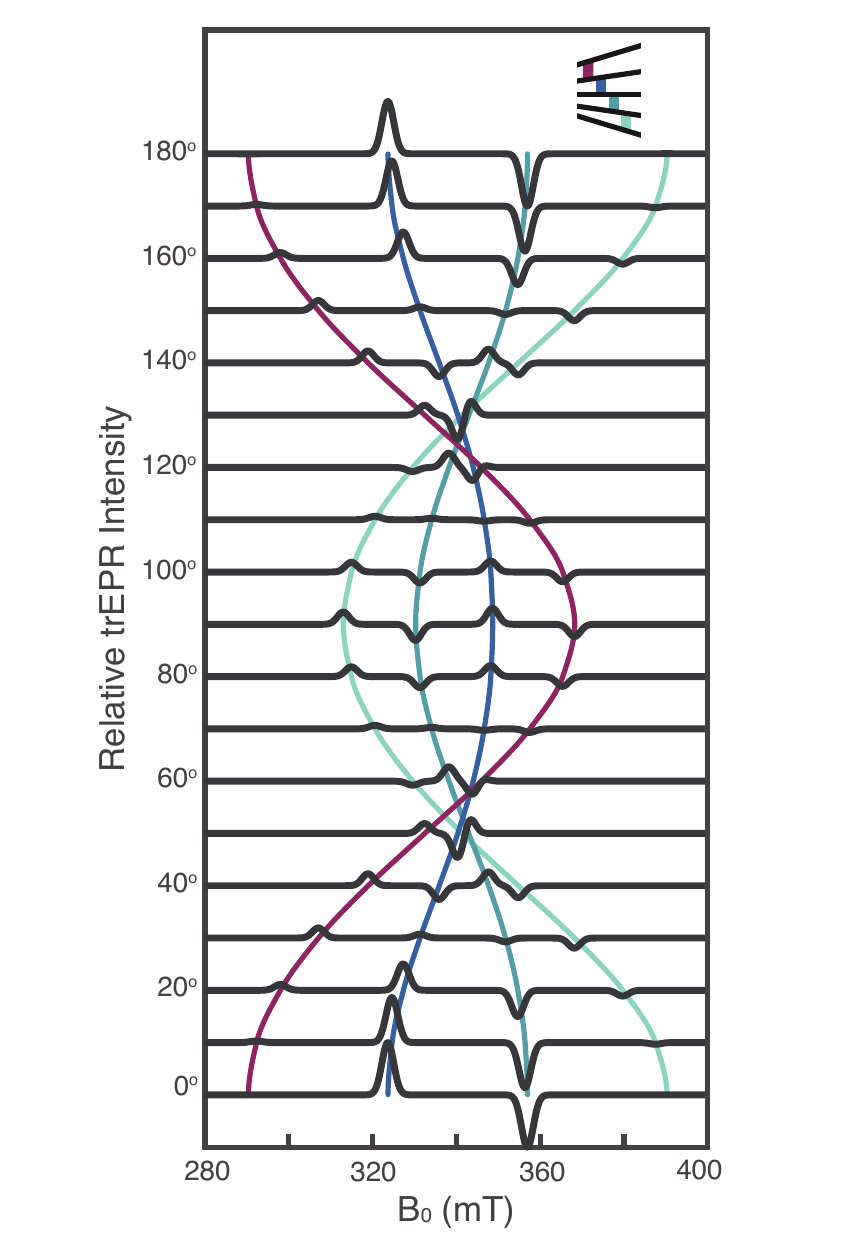}
    \caption{ Calculated prompt EPR spectra for the rotation of an oriented and parallel sample with respect to an applied field. $\theta$ is the polar angle between the lab-fixed Zeeman axis and the principal $z'$-axis. Peak intensities are proportional to differences in state populations and are relative to intensities at $\theta = 0^{\circ}$. Differences in the number and sign of the peaks are the result a $\theta$-dependent coupling between the $|^1\text{TT}\rangle$ and the adiabatic $|\alpha\rangle$ states described in the text. Colored lines follow specific transitions, indicated in the inset where states are ordered in energy with respect to the applied field. Parameters for the calculation are consistent with those reported in Fig.~\ref{fgr:FIG5}.}
    \label{fgr:FIG4}
\end{figure}

Changing the orientation of the dimer relative to the Zeeman field results in a nonperturbative change in the hamiltonian. The transitions from $^1\text{TT}$ to the $^5\text{TT}_{\pm1}$ sublevels, that were once symmetry forbidden, are now allowed and state selectivity diminishes. For dimers in a powder or frozen solution, the EPR signal is a sum over an ensemble of molecules that have a broad distribution of orientations with respect to the Zeeman field. The resulting spin polarization is scrambled, which leads to decoherence in the ensemble signal. In quantum information applications this is a source of noise, and it is therefore important not only to fix the molecular axes relative to one another, but also with respect to the laboratory axis.

For dimers with their principal axes fixed in space, the prompt EPR spectra as a function of the polar angle, $\theta$, between $z$ and $z'$, exhibit different numbers of peaks with different frequencies and signed relative intensities (Fig.~\ref{fgr:FIG4}). The coupling term, $V$, is a function of $\theta$, and all quintet states may be directly accessible from the $^1\text{TT}$ state. There is intensity-borrowing from the $^5\text{TT}_0$ and $^5\text{TT}_{\pm2}$ zero-field states into all $^5\text{TT}_{M}$ sublevels as a function of $\theta$. As $\theta$ goes from 0$^\circ$ to $90^\circ$, state selectivity for the relaxation from $^1\text{TT}$ monotonically decreases. The EPR signal is a much weaker function of the azimuthal angle, $\phi$.

\begin{figure}[H]
    \centering
    \includegraphics{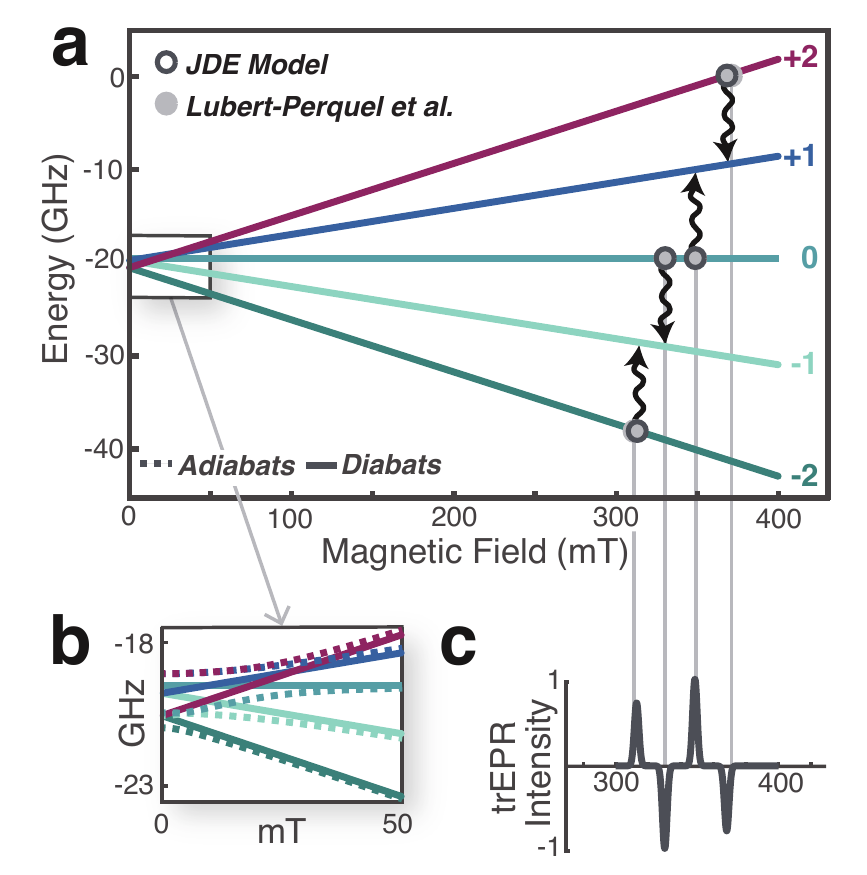}
    \caption{The $\theta = 90^\circ$ field-swept prompt EPR quintet spectrum for a parallel dimer of pentacene molecules in a dilute crystal. \textbf{a} The energies of diabatic quintet states as a function of the Zeeman field. Parameters used are literature values for parallel pentacene molecules in \textit{p}-terphenyl \protect{\cite{Lubert-Perquel2018}}. The experiments performed in Ref. \protect{\citenum{Lubert-Perquel2018}} are in the strong field limit, where the transitions are far removed from the avoided crossings between adiabatic states (\textbf{b}). In this limit, the diabatic states (solid lines) of the Zeeman hamiltonian are very close to the adiabatic states (dashed lines). Open circles indicate predicted transitions for the excitation frequency 9.538 MHz, which compare favorably to the field values of transitions reported by Lubert-Perquel et al. (closed circles)\protect{\cite{Lubert-Perquel2018}}. At this field-orientation, the singlet state population transfers to the quintet $M=0,\pm2$ states. Wavy arrows indicate the direction of transitions for $D>0$. \textbf{c} Simulated spectrum, where positive changes in the EPR intensity indicate induced absorption and negative changes indicate stimulated emission.  The relative peak intensities were calculated from coupling matrix elements between the singlet and quintet states using a short-time approximation to the Pauli master equation. The sign of $E$ dictates the relative peak magnitudes, and $E<0$ produces strong inner peaks with slightly weaker outer peaks, which is consistent with the spectrum reported in Ref. 14. Field positions of their spectrum for the herringbone dimer geometry are indicated by vertical lines and are in excellent agreement with our calculations for the parallel dimer.}
    \label{fgr:FIG5}
\end{figure}

Lubert-Perquel et al. published $\text{TT}$ EPR spectra for dilute crystals of pentacene molecules doped into a \textit{p}-terphenyl matrix \cite{Lubert-Perquel2018}. The pentacene molecules adopted both parallel and herringbone configurations in the host, as observed in the pentacene crystal structure. This elegant design allowed them to disperse the dimers and fix their orientations in space. The $^5\text{TT}$ EPR spectrum contains contributions from dimers in both parallel and herringbone geometries. Figure~\ref{fgr:FIG5}a shows the field-swept energies of $^5\text{TT}_M$ states for parallel chromophores, where the Zeeman field, $B_0$, is directed along the principal $x'$-axis, which is the most well-resolved spectrum in their paper. On this diagram, we indicate the diabats which are pure $|{S,M}\rangle$ states, along with the adiabats described above (Fig.~\ref{fgr:FIG5}b). The transitions occur in the strong field limit, far from the avoided crossings between the adiabats, which occur at much smaller values of the $B_0$ field. Indeed, in the vicinity in which the EPR transitions are recorded, the diabats and adiabats very nearly coincide. 

We present our calculation for the spectrum in Fig.~\ref{fgr:FIG5}c. The parameters used to generate the spectra were taken from Ref. 14, and as illustrated in Fig.~\ref{fgr:FIG5}, our model fits their data extremely well, with one caveat. The polarization pattern (AEAE) present in the calculated spectrum for parallel molecules (Fig.~\ref{fgr:FIG5}c) matches the spectrum they assigned to the herringbone configuration. A similar calculation of the EPR spectrum for herringbone dimers was carried out with the (non-parallel) \emph{JDE} model and matches the spectrum that Lubert-Perquel et al. had assigned to parallel dimers. We would thus re-assign their spectra; their EEAA spectrum is, rather, attributed to the herringbone configuration and their AEAE spectrum to the parallel configuration.

\section*{Discussion} 

We have provided a derivation and an analysis of a model hamiltonian for singlet fission with an eye toward quantum computing, information, and sensing applications. The model is specified by only three parameters: the inter-chromophore isotropic exchange coupling, $J$, and the intra-chromophore ZFS axial parameter, $D$, and rhombicity parameter, $E$. These parameters can be measured independently or calculated using electronic structure. The model one arrives at under a set of reasonable approximations is something we call the \emph{JDE} model, named for the $J$, $D$ and $E$ parameters of that hamiltonian.
 
 In particular, we have shown that one can use the magnetic sublevels  of the $^5\text{TT}$ space as ``qudits'' in quantum information applications\cite{Brennen2005}, where EPR experiments perform the function of quantum gates. The five quintet states offer a quantum advantage over the three states of the spin-polarized triplets, produced either by intersystem crossing or as the final spin unpaired products of a singlet fission process in a crystal. We have shown the conditions under which the $^1\text{TT}$ state transfers to states in the quintet block and have given the conditions for maximal state selectivity, and thereby the most efficient pathway to optical spin polarization for those transitions. 
 
To decrease the transition rates from the $^{2S+1}\text{TT}_M$ manifold into the incoherent unpaired triplets, one needs to keep the value of $J$ large. This implies that molecular dimers that are covalently bound to one another or doped as an impurity component into a host crystal are ideal candidates for generating optically spin-polarized quantum states near room temperature.

We have identified, for the first time, the selection rules for relaxation between the various doubly excited $\text{TT}$ levels in chromophores with parallel symmetry at both zero and large Zeeman fields. At zero field, fluctuations in $J$ transfer population from the $^1\text{TT}$ state into the maximally entangled quintet state, $Q_{z^2}$. This transition rate goes as $D^2$. There is one and only one other allowed transition, which is into the $Q_{x^2-y^2}$ state, but this rate is proportional to $E^2$ and is much slower. One can make the relaxation even more state selective by synthesizing molecules with large $|D|/|E|$ ratios.

In the strong field conditions, characteristic of both time-resolved field swept EPR experiments and quantum computing applications, we find that relaxation can be kept state-selective provided that the principal axes of the two chromophores are parallel to each other and to the Zeeman field. When a molecule's principal $z'$-axis does not align with the Zeeman axis, several symmetries are broken, and transitions are possible to all sublevels in the quintet block. In samples where the molecules have a broad distribution of orientations relative to the Zeeman axis, the ensemble will exhibit decoherence. This is a different source of decoherence than, for example, inhomogeneous broadening, that can be removed through echo techniques. This means that one needs to also devise a method to immobilize and control the orientation of the singlet fission chromophores. Recent, elegant, experimental work has shown that this is possible\cite{Lubert-Perquel2018}.

Finally, using our model and analysis, we calculated the prompt EPR spectra of pentacene dimers doped into a \textit{p}-terphenyl crystal. With the parameters $J$, $D$ and $E$ provided, there are no adjustable parameters in this calculation. The fit to the experimental data is very good, but where Lubert-Perquel et al. would assign the spectral component to the parallel geometry of a dimer, we assign it to the herringbone (Fig. \ref{fgr:FIG5}) \cite{Lubert-Perquel2018}.
  
Singlet fission can create strongly spin-polarized products and thereby generate nearly pure quantum states at room temperature, but there are several design principles that one should follow. First, keep the inter-chromophore exchange, $J$, large. Second, immobilize the molecules and align their principal axis to the Zeeman field. Both requirements are satisfied in immobilized and oriented, covalently linked dimers, or in dimer pairs that are embedded in a crystal host that inhibits exciton diffusion\cite{Lubert-Perquel2018}.

Singlet fission can offer many of the quantum advantages found in color centers, like the NV center in diamond, but with a bottom-up approach to design that is currently unavailable in color centers whose defects are implanted in the material post-synthesis. Singlet fission, in contrast, is able to capitalize on the arsenal of synthetic techniques developed in organic chemistry to design molecules. Our work provides a quantitative model for computing dynamics and fitting spectra, and qualitative design principles for the synthetic design of new organic molecules for quantum information applications. This is an important step in establishing the relationship between molecular structure and function in an emerging class of organic, novel quantum materials.

\subsection*{Acknowledgements}
We thank Obadiah Reid and Justin Johnson for useful discussions regarding magnetic resonance experiments. We thank Niels Damrauer for many useful and insightful discussions. Funding was provided by the United States Department of Energy, Office of Basic Energy Sciences (ERW7404).

\subsection*{Author contributions}
JE designed research. JE and KS performed research, analyzed data and wrote the manuscript.

\subsection*{Competing interests} 
The authors declare no competing interests.

\subsection*{Data Availability Statement}
The data supporting the findings of this study are available from the corresponding author upon request.

\end{document}